\newif\ifoverleaf

\ifoverleaf
\documentclass[twocolumn]{article}
\usepackage[margin=1in]{geometry}
\else
\documentclass[10pt,conference]{IEEEtran}
\fi

\usepackage{amsmath,amsfonts,amssymb}
\usepackage{graphicx} 
\usepackage{braket}
\usepackage{dsfont}
\usepackage{color}
\usepackage{xspace}

\newcommand{\mtx}[1]{\boldsymbol{#1}}
\renewcommand{\vec}[1]{\boldsymbol{#1}}

\title{Gauging practical computational advantage using a classical, threshold-based Gaussian boson sampler}
\author{Sarvesh Raghuraman, Aditya Patwardhan, Brian La Cour}
\date{March 2025}

\ifoverleaf
\author{Sarvesh Raghuraman, Aditya Patwardhan, Brian La Cour}
\else
\author{
\IEEEauthorblockN{Sarvesh Raghuraman}
\IEEEauthorblockA{\textit{Applied Research Laboratories} \\
\textit{The University of Texas at Austin}\\
Austin, Texas, United States \\
sarvesh.raghuraman@arlut.utexas.edu}
\and
\IEEEauthorblockN{Aditya Patwardhan}
\IEEEauthorblockA{\textit{The University of Texas at Austin}\\
Austin, Texas, United States \\
apatwardhan@utexas.edu}
\and
\IEEEauthorblockN{Brian La Cour}
\IEEEauthorblockA{\textit{Applied Research Laboratories} \\
\textit{The University of Texas at Austin}\\
Austin, Texas, United States \\
blacour@arlut.utexas.edu}
}
\fi

\begin{document}

\maketitle

\begin{abstract}

We describe an efficient, scalable Gaussian boson sampler based on a classical description of squeezed quantum light and a deterministic model of single-photon detectors that ``click'' when the incident amplitude falls above a given threshold.  Using this model, we map several NP-Complete graph theoretic problems to equivalent Gaussian boson sampling problems and numerically explore the practical efficacy of our approach.  Specifically, for a given weighted, undirected graph we examined finding the densest $k$-subgraph and the maximum weighted clique.  We also examined the graph classification problem.  Compared to traditional classical solvers, we found that our method provides better solutions in a comparable amount of samples for graphs with up to 2000 nodes. 
\end{abstract}



\section{Introduction}

Over the last decade, the field of quantum computation has grown enormously, driven in part by significant advances in hardware design and system control.  Tasks such as random circuit sampling, prime factorization, and unstructured searches have all been identified as possible means for demonstrating quantum advantage.  Photonic quantum computers have emerged as a dark horse in the race that promises to provide a fast-track to a scalable, fault-tolerant quantum computer.  Along the way, Boson sampling has emerged as a practical method for demonstrating quantum advantage.

The Boson sampling problem can be described simply as follows: given a particular input Fock state of $M$ photonic modes that are input to a linear circuit with $M$ output modes, represented by an $M \times M$ unitary $\mtx{U}$, one seeks to determine the distribution of output photons.  The problem is provably \#P-hard and may be solved by either producing the complete probability distribution or by some mechanism that faithfully produces samples from this distribution \cite{Aaronson2013}.  Unlike traditional, gate-based quantum computers, Boson samplers do not provide a universal model of quantum computation, but they can be used as a means for demonstrating quantum advantage.

A variant of this approach, Gaussian boson sampling (GBS), uses squeezed light instead of Fock states \cite{Hamilton2017}.  GBS has the same computational complexity as Boson sampling, but is more amenable to experimental implementation, with several recent experiments claiming to have demonstrated quantum computational advantage \cite{Zhong2020,Zhong2021,Deng2023}.  Recently, it has been found that Gaussian boson samplers can be harnessed to encode and solve several problems of practical interest.  Researchers at Xanadu have explored a variety of such problems that could achieve speedups over purely classical methods using a GBS device, most being of the graph theoretic variety \cite{Bromley2020}.  This opens the exciting possibility that a Gaussian boson sampler could be used not only for demonstrating quantum advantage, but for solving practical, real-world problems.  A practical challenge is that current GBS devices are cumbersome, expensive, and difficult to operate.  A typical GBS device involves multiple sources of non-classical light, a complex linear optical interferometer to encode the problem, and an array of high-efficiency detectors, all delicately aligned for phase coherence.

In this paper, we offer a simple, practical alternative that can be implemented on a classical digital computer yet still achieve the same theoretical advantage over traditional solvers.  Specifically, we introduce an efficient, scalable Gaussian boson sampler based on a classical description of squeezed quantum light and a deterministic model of single-photon detectors that ``click'' when the incident amplitude falls above a given threshold.  We examine the performance of this Threshold-based Gaussian boson sampler (TGBS) in solving several graph-theoretic problems and compare it against more traditional algorithms.  One may also consider the performance of our TGBS as a benchmark against which to compare future photonic GBS devices.

The organization of the paper is as follows: In Sec.\ \ref{sec:GBS} we discuss the basic concepts of GBS, introduce our classical TGBS model, and describe how to map certain graph problems into the GBS framework.  Section \ref{sec:NumSim} describes our application of TGBS to the problems of finding the densest $k$-subgraph, identifying the maximum cliques (weighted and unweighted), and classifying graphs as proposed by Bromley et al \cite{Bromley2020}.  Our conclusions are summarized in Sec.\ \ref{sec:Con}.


\section{Gaussian Boson Sampling}
\label{sec:GBS}


\subsection{Threshold-based Gaussian Boson Sampling}

Let $M$ be the number of input modes into the GBS device, and let $r_m$ and $\phi_m$ denote the squeezing strength and phase for mode $m \in \{1, \ldots, M\}$.  The quantum mechanical description of the initial squeezed vacuum state $\ket{r_m,\phi_m}$ is given by a separable state defined by the transformed annihilation operator $(\cosh r_m) \hat{a}_m + e^{i\phi_m} (\sinh r_m) \hat{a}_m^\dagger$ obtained via the Bogoliubov transformation, where $\hat{a}_m^\dagger$ and $\hat{a}_m$ are the creation and annihilation operators, respectively, for mode $m$ \cite{Agarwal2012}. 

In our classical model, the annihilation operator $\hat{a}_m$ is replaced by the random variable $\sigma z_m$, where $\sigma^2 = \frac{1}{2}$ represents the energy per vacuum mode (in units of $\hbar\omega$) and $z_m$ is a standard complex Gaussian random variable (i.e., with $\mathsf{E}[z_m] = 0$, $\mathsf{E}[|z_m|^2] = 1$, and $\mathsf{E}[z_m^2] = 0$).  This matches the statistical distribution of the vacuum state according to quantum electrodynamics \cite{Milonni1993}.  All $z_1, \ldots, z_M$ are independent and identically distributed.  The squeezed state amplitude for input mode $m$ is then modeled as
\begin{equation}
a_m = (\cosh r_m) \sigma z_m + e^{i\phi_m} (\sinh r_m) \sigma z_m^* \; ,
\end{equation}
which, for $r_m > 0$, is an improper complex Gaussian random variable with $\mathsf{E}[a_m] = 0$, $\mathsf{E}[|a_m|^2]  = (\sinh r_m)^2 + \frac{1}{2}$, and $\mathsf{E}[a_m^2] = e^{i\phi_m} (\cosh r_m) (\sinh r_m)$.  These correspond to the expectation values of the operators $\hat{a}_m$, $(\hat{a}_m^\dagger \hat{a}_m + \hat{a}_m \hat{a}_m^\dagger)/2$, and $\hat{a}_m^2$, respectively, with respect to the squeezed state $\ket{r_m,\phi_m}$.  More generally, for any symmetrically ordered operator the quantum expectation and corresponding classical expectation will match exactly, by the optical equivalence principle \cite{Sudarshan1963,Cahill1969a,Cahill1969b}.  Note that the average photon number is given by $\overline{n} = (\sinh r_m)^2 = \mathsf{E}[|a_m|^2] - \frac{1}{2}$.

Although the agreement with expectations of symmetrically ordered operators is exact, deviations may arise when considering discrete measurement outcomes.  A novelty of our classical model, indeed the only novelty, is that a detection on mode $m$ is said to occur when $|a_m| > \gamma_m$ for a given threshold $\gamma_m \ge 0$.  This threshold-based detection scheme has been described elsewhere as a means for approximating the Born rule \cite{LaCour2020}.  Letting $\Pr[\cdot]$ denote the probability measure over the random variables, the classical probability of a detection on input mode $m$ is then given by $\Pr[ |a_m| > \gamma_m]$.  This allows us to map continuous amplitudes to discrete (binary) outcomes.  Note that dark counts and missed detections are still possible within this framework, and their frequency will depend implicitly on the threshold value, $\gamma_m$, chosen for each mode.  Typically, this is chosen to be of order unity, but it is a design choice that may affect the speed and accuracy of the algorithm.

For Gaussian boson sampling we consider an $M \times M$ unitary matrix $\mtx{U}$ representing an optical interferometer with an equal number, $M$, of input and output states.  If some input modes are not used, we may set their squeezing strength to zero to represent vacuum states.  If the output modes feed into pseudo photon number resolving detectors, these, too, may be incorporated into $\mtx{U}$, with the number of modes suitably increased.  Thus, without loss of generality we shall assume an equal number of input and output modes with threshold detectors (in the classical or quantum sense) at each output.

Application of the unitary is straightforward.  Let $\vec{a} = [a_1, \ldots, a_M]^\mathsf{T}$ denote the vector of random amplitudes representing the $M$ modes of the input squeezed states.  The output vector is simply
\begin{equation}
\vec{a}' = \mtx{U} \vec{a} \; .
\end{equation}
The covariance, $\mtx{U} \mathsf{E}[\vec{a} \vec{a}^\dagger] \mtx{U}^\dagger$, of $\vec{a}'$ may be non-diagonal, unlike that of $\vec{a}$, and the pseudo covariance, $\mtx{U} \mathsf{E}[\vec{a}\vec{a}^\mathsf{T}] \mtx{U}^\mathsf{T}$, may be nonzero.  Although $\vec{a}'$ is a Gaussian random vector, it will in general be both correlated and improper, both key properties of a multi-mode entangled state \cite{LaCour2021}.  As with $\vec{a}$, the classical expectation values of functions of $\vec{a}'$ will match exactly the quantum expectation values for corresponding symmetrized operators.  In this sense, $\vec{a}'$ is an exact representation of the Wigner function for the corresponding transformed quantum state.  Using the aforementioned threshold detection scheme, these amplitudes will be mapped to binary values to produce binary string samples.

\subsection{GBS Programming of Undirected Graph Problems}

For a classical digital computer, operations are carried out on a set of binary input states through a sequence of logical gates.  For a gate-based quantum computer, operations are performed on a quantum state via a sequence of unitary gates. In the case of GBS, the interferometer, represented by the unitary matrix $\mtx{U}$, operates on a set of input squeezed vacuum states, here represented as a set of complex Gaussian random vectors. Thus, to program a graph problem into a GBS device, we need to formulate a unitary matrix and a set of squeezing strengths from the defined graph problem.

An undirected graph is defined by a set of nodes and connecting edges, along with any associated weights.  The edges of an undirected graph define a symmetric adjacency matrix $\mtx{A}$.  Using the Autonne–Takagi decomposition, such matrices may be factored in terms of a unitary matrix $\mtx{U}$ and a real, nonnegative diagonal matrix $\mtx{\Lambda}$ such that \cite{Houde2024}
\begin{equation}
    \mtx{A} = \mtx{U} \mtx{\Lambda} \mtx{U}^\mathsf{T}
    \label{eq:takagi-autonne-decomposition}
\end{equation}
where $\mtx{\Lambda} = \mathrm{diag}(\lambda_1, \ldots, \lambda_M)$ and $M$ is the number of nodes.

In a GBS experiment, $\mtx{U}$ defines the interferometer, while the diagonal elements of $\mtx{\Lambda}$ define the squeezing strengths, $r_1, \ldots, r_M$, for each of the $M$ input modes.  Specifically,
\begin{equation}
    r_m = \tanh^{-1}(\lambda_m) \; ,
\end{equation}
and the mean photon number, $\overline{n}$, is therefore
\begin{equation}
  \overline{n} = \sum_{m=1}^{M} (\sinh r_m)^2 = \sum_{m=1}^{M} \frac{\lambda_m^2}{1-\lambda_m^2} \; .
\end{equation}
Optionally, the squeezing parameters can be rescaled for a desired mean photon number.  For simplicity, the phase, $\phi_m$, is taken to be zero.  To perform this encoding, we use the graph embedding function provided by Xanadu's Strawberry Fields package and from the previous definitions, we can fully describe a GBS experiment corresponding to a given undirected graph \cite{Bromley2020}.

An exact classical implementation of this GBS experiment would require computation of the probabilities for all $2^M$ possible binary strings of length $M$, which in turn requires computation of a Torontonian for each such string \cite{Quesada2018}.  This is a problem of complexity \#P and hence, is classically intractable \cite{Quesada2018}.  A physical, quantum implementation would require a complex, expensive, and easily configurable optical interferometer, which we do not consider.  Instead, we examine the use of a threshold-based Gaussian boson sampler, as described earlier.  This is an inexact classical scheme meant to mimic the behavior and performance of real, albeit nonideal, optical GBS experiments.  In this work, we shall not be concerned with whether this approach truly mimics the performance and higher-order correlations of a real, optical GBS.  Rather, we shall examine its performance in comparison to standard classical seeded algorithms, as discussed in Sec.\ \ref{sec:NumSim}.


\section{Numerical Simulations}
\label{sec:NumSim}

As a classical scheme, it is important to understand the scaling complexity of the TGBS algorithm, for which there are 3 stages to consider.  There is a front-end cost to perform the Autonne–Takagi decomposition, which has a time complexity of $\mathcal{O}(M^3)$ \cite{LAPACK2008}. Next, there is the generation of the initial random amplitudes $a_{m,n}$ for each $m \in \{1, \ldots, M\}$ mode and $n \in \{1, \ldots, N\}$ realization, which has a time complexity of $\mathcal{O}(MN)$.  Finally, matrix multiplication between the $M \times M$ unitary matrix and $M \times N$ matrix of amplitudes will have a time complexity of $\mathcal{O}(M^2 N)$.  Since we typically take $N \ll M$ to be a constant (primarily for the graph searching problems), the initial Autonne–Takagi decomposition is expected to be the dominant scaling factor.  For a given graph problem, each of the $N$ realizations provides an initial subgraph (identified by modes with an outcome of one) that is used as input to a standard classical algorithm.  Our hypothesis is that this initial seed provides faster convergence to a more accurate solution than the traditional approach of using randomly chosen seeds.

To generate random graphs for our investigation, we used the Erd\"{o}s–R\'{e}nyi model, which defines a graph by a set of nodes and a given probability, $p$, of an edge existing between any pair of nodes \cite{Erdos1959}. Since we aim to benchmark the GBS algorithm on its scalability in the context of identifying dense, connected components, we set $p = \log(n)/n$, where $n$ is the number of nodes in the graph. This edge probability is known to create a graph with a sharp threshold in connectedness as $n$ tends to infinity. Note that the expected number of edges for such a graph is given by $\binom{n}{2} \log(n)/n$ and so, scales as $n \log(n)$.  For our experiment, we took the average score over 100 random graphs for each particular size in order to obtain statistically meaningful results.
Lastly, for all of our simulations, we used Python on a 13th Gen Intel\textregistered \xspace Core\texttrademark \xspace i5-13400, 2500 Mhz, 10 Core processor and 16 GB of RAM.

\begin{figure}[ht]
\includegraphics[width=8cm]{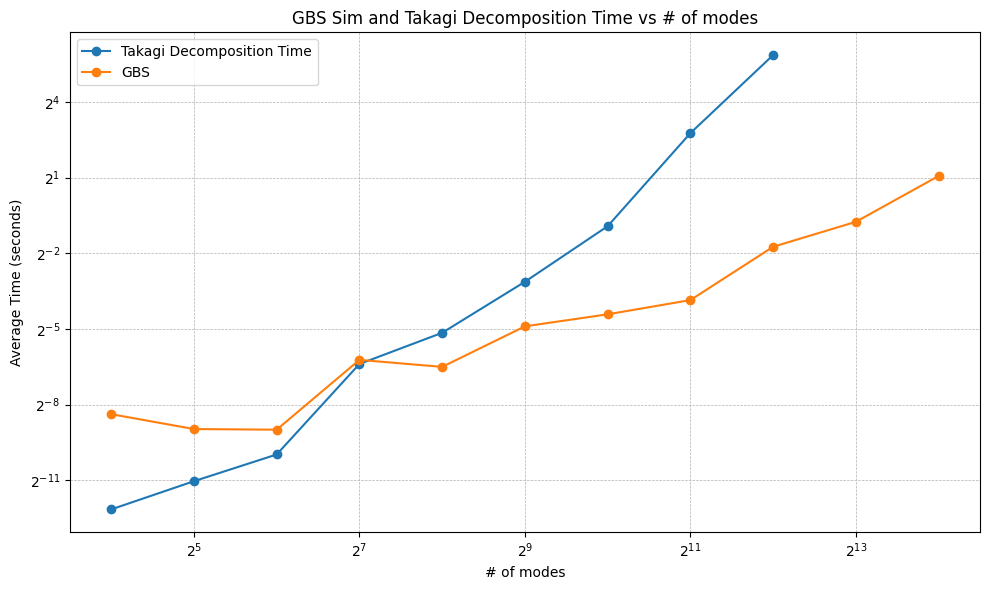}
\caption{Runtimes of the Autonne–Takagi decomposition (blue) and TGBS simulation (orange), averaged over 100 random graphs for each size.}
\label{fig:gbsruntime}
\end{figure}

Figure~\ref{fig:gbsruntime} shows the time to perform the Autonne–Takagi decomposition for random graphs from $2^4$ to $2^{14}$ nodes.  This is compared against the time to perform the TGBS simulation itself and here, we find that the decomposition indeed dominates the computation time and scales as expected. 

\subsection{Subgraph Density}

A key feature of GBS is its ability to sample subgraphs of high density.  To test the quality of the initial seed graphs, we generated random graphs of size $M \in \{2^4, 2^5, \ldots, 2^{14}\}$ and performed TBGS to produce 20 subgraphs. For a subgraph of $k \le M$ nodes, we measured its quality by the density, $\rho$, defined as the ratio of the number of edges, $\varepsilon$, to the number of possible edges, $k(k-1)/2$.  Thus,
\begin{equation}
    \rho = \frac{2 \varepsilon}{k(k - 1)} \; .
    \label{eq: subgraph density}
\end{equation}

\begin{figure}[ht]
\includegraphics[width=8cm]{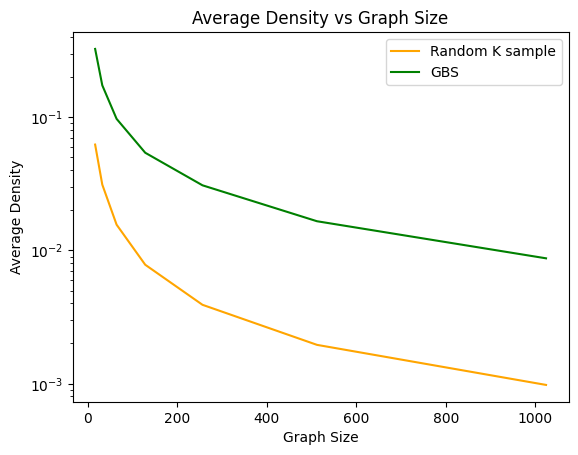}
\caption{Plot of average density of seeds obtained through GBS and random methods}
\label{fig:seeddensity}
\end{figure}

For a baseline comparison, we also found the mean density of a subgraph of $k$ randomly chosen nodes. To do this, we computed the mean subgraph size, $\bar{j}$, generated by TGBS for each initial graph size. Then, for each original graph, we gave each node a probability of $1/\bar{j}$ to be included in the subgraph. We observe in Figure~\ref{fig:seeddensity} that GBS does indeed offer a higher quality seed than that obtained by random sampling. Note this metric does not consider node or edge weights, simply the nodes and edges themselves. In what follows, we investigate whether these higher density seeds actually result in faster convergence and higher quality solutions.


\subsection{Densest $k$-Subgraph}
Given our density criterion, a natural problem to consider is that of finding the densest $k$-subgraph. For a given undirected graph, the task here is to find the densest connected subgraph of $k$ nodes.  This problem is in the class of NP-Hard problems, meaning that finding a solution is believed to be classically intractable. As stated previously, GBS will sample from dense regions of the graph, giving us a dense seed subgraph with high probability and then search from there. We will not include disjoint subgraphs, which is indicated by these disjoint graphs being ``pruned.''

For the experiment, we implemented this seed-and-search process across different types of seed graphs as follows:
\begin{enumerate}
    \item GBS sample: The seed graph is provided by a sample from a programmed GBS simulation.
    \item Random Single Node sample: The seed graph is a randomly selected node from the graph.\
    \item Random J Node Sample: Randomly sample $J$ nodes from the graph, where $J$ is the mean size of the subgraph sampled by GBS
    \item Greedy Peeling: The seed graph is the entire graph
\end{enumerate}

For a given graph, our process for finding the densest $k$-subgraph is based on the algorithm by Charikar, and is defined as follows \cite{Charikar2000}: 
\begin{enumerate}
    \item For each seed graph perform the following loop:
    \begin{enumerate}
        \item If the size of the current subgraph is less than $k$, include the node that is adjacent to the current subgraph with maximum degree. 
        \item If the size of the current subgraph is greater than $k$, remove the node of least degree.
        \item If the size of the current subgraph is equal to $k$, exit the loop and return the current subgraph.
    \end{enumerate}
\end{enumerate}
Note that ties are broken arbitrarily.

Lastly, we define our testing graphs as planted subgraphs consisting of two connected Erd\"{o}s–R\'{e}nyi graphs. In this system, two Erd\"{o}s–R\'{e}nyi graphs, $\alpha$ and $\beta$, are generated, with edge probabilities $p_1 \gg p_2$, respectively. The two graphs have edges assigned between random pairs of nodes with probability $p_2$. We arbitrarily set $p_1 = 0.75$ and $p_2 = 0.1$. Additionally, the sizes of the dense inner and less dense outer graphs have a 1:9 ratio.

\begin{figure}[ht]
\includegraphics[width=8cm]{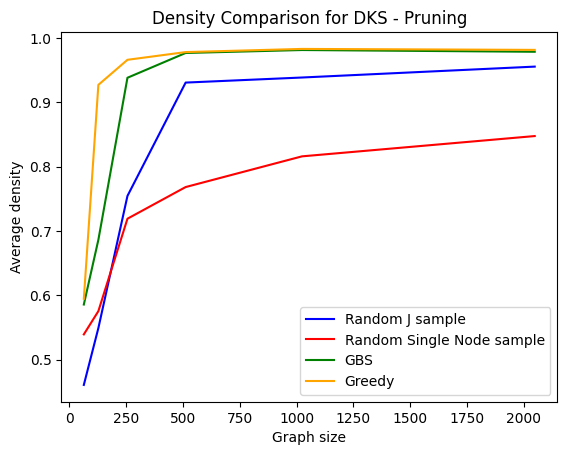}
\centering
\caption{Plot of average density versus graph size for planted solutions.}
\label{fig:plantedDKSdensity}
\end{figure}

\begin{figure}[ht]
\includegraphics[width=8cm]{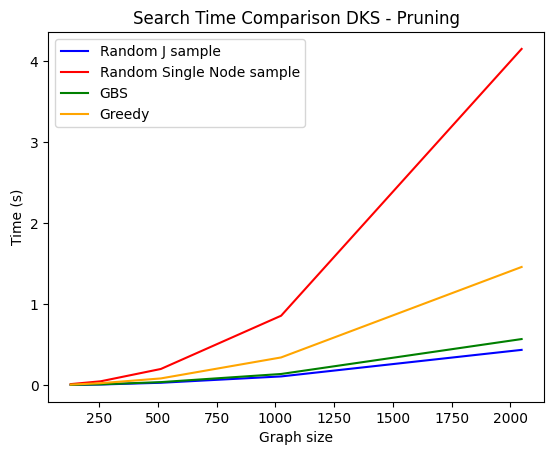}
\caption{Plot of time versus graph size for planted solutions.}
\centering
\label{fig:plantedDKSinitsearchtime}
\end{figure}

We observe in Figure~\ref{fig:plantedDKSdensity} that the GBS algorithm consistently delivers a high density solution in the planted graphs, alongside the greedy strategy. Additionally, we note from Figure~\ref{fig:plantedDKSinitsearchtime} that the GBS seed's searchtime is significantly less than the greedy strategy as well as the other strategies, indicating a faster convergence rate. In comparison to the Random J seed, we witness the GBS seed taking slightly longer, but providing a higher quality solution as well. From these results, we can confirm that the GBS seed does indeed provide an advantage over classical seed based methods for finding dense subgraphs.


\subsection{Maximum Clique/Maximum Weighted Clique}

While the densest $k$-subgraph problem imposes a constraint on the number of nodes and maximizes density, one could instead constrain the density and aim to maximize the number of nodes.  In the most extreme case, the subgraph is a clique and, thus, has a density of one.  This gives rise to the maximum clique problem: given an undirected graph, the task is to find the clique with the maximum possible number of nodes.    

Similar to the densest $k$-subgraph problem, the most prominent algorithms for identifying maximum cliques are those which expand and shrink from a given seed (usually random) \cite{Bromley2020}. We aim to see if a GBS derived seed can provide a potential improvement. We employ a method similar to that of the previous problem with a seeded subgraph obtained from the GBS device. From this, we use the algorithm detailed by Bromley \textit{et al.}, which shrinks the seed graph until a clique is found, by repeatedly removing the node of lowest degree \cite{Bromley2020}. This is followed by multiple cycles of growing and swapping. A node outside the current clique, but connected to every node in it, is added to the clique at each step of the growing stage. When no more such nodes exist, the swapping stage begins. In the swapping stage, a randomly selected node of the clique is swapped for a node outside the clique such that the resulting graph remains a clique, after which the growth stage occurs. 

We conduct our tests on the same construction of random planted graphs as the densest $k$-problem, and compare our results against the other seeded strategies from before.

\begin{figure}[ht]
\includegraphics[width=8cm]{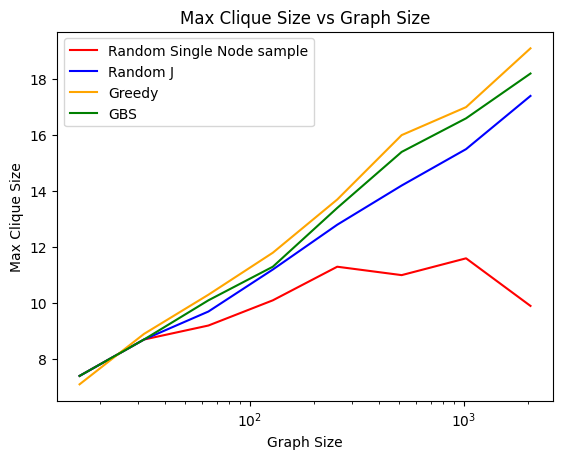}
\caption{Plot of clique size versus graph size.}
\label{fig:maxcliqueovergraphsize}
\end{figure}

\begin{figure}[ht]
\includegraphics[width=8cm]{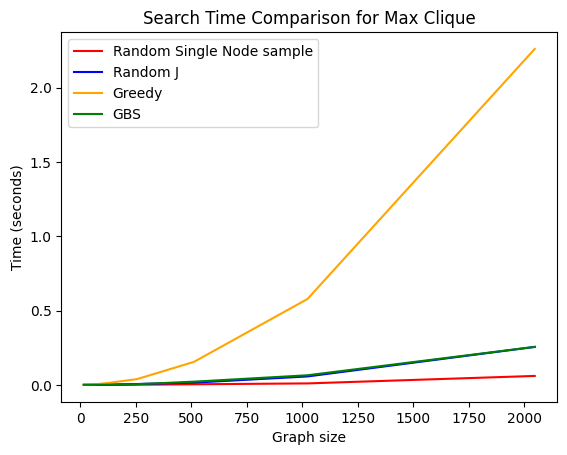}
\caption{Plot of time versus graph size for the max clique problem.}
\label{fig:maxcliqueinittime}
\end{figure}

In Figure \ref{fig:maxcliqueovergraphsize} we witness the GBS strategy producing consistently higher quality solutions over the current maximum clique strategies referenced in Bromley \textit{et al.}. However, we do note the slight increase in time from the random seed strategies in Figure \ref{fig:maxcliqueinittime}. This results in a tradeoff between solution quality and time prompting  additional probing to demonstrate a definitive difference between GBS and the Random J strategies. For this, we look to a related and more involved problem.

A generalized version of the previous problem is that of finding the weighted maximum clique, where the goal is to find a clique in a node-weighted graph which has the maximum sum of node weights. In other words, the maximum clique problem can be thought of as an instance of maximum weighted clique with uniform node weights.  Once again, a seed graph is sampled from the GBS device. However, the graph must be encoded differently from the maximum clique and densest $k$-subgraph problems, so that the sampled graphs are both dense and have a high weight. To do this, the adjacency matrix is transformed before programming the GBS device using the formula $A' = \Omega(D-A)\Omega$,\cite{Banchi2020} where $D - A$ is the graph Laplacian and $\Omega$ is a diagonal matrix defined as 1 + $\alpha w_{ii}$ where $\alpha$ is the constant which encodes the relative importance of the weights $w$. Afterwards, the matrix $U$ is derived as usual and the algorithm continues as with the maximum clique problem. Lastly, for the following tests, we assign a uniformly random weight between 0 and 1 to each node.

\begin{figure}[ht]
\includegraphics[width=8cm]{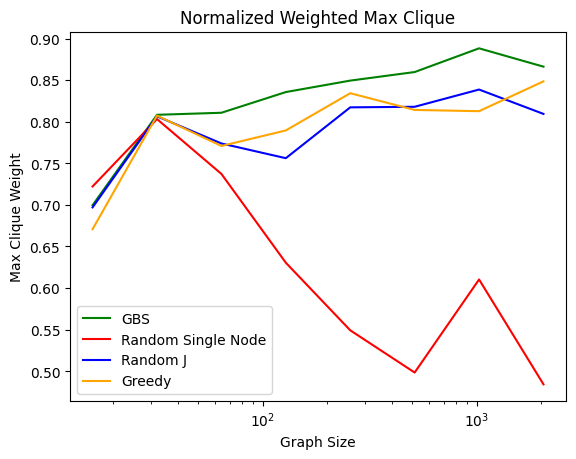}
\caption{Plot of clique weight versus graph size for the weighted maximum clique problem.}
\label{fig:wmdormalizedlogscale}
\end{figure}

\begin{figure}[ht]
\includegraphics[width=8cm]{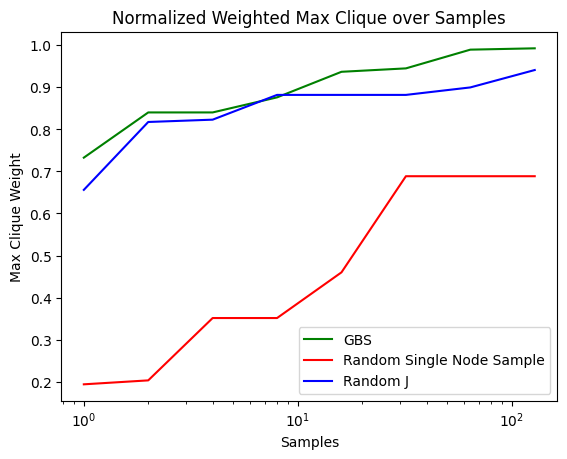}
\caption{Plot of clique weight versus graph size up to 2048 for the weighted maximum clique problem.}
\label{fig:WMCsamples}
\end{figure}

\begin{figure}[ht]
\includegraphics[width=8cm]{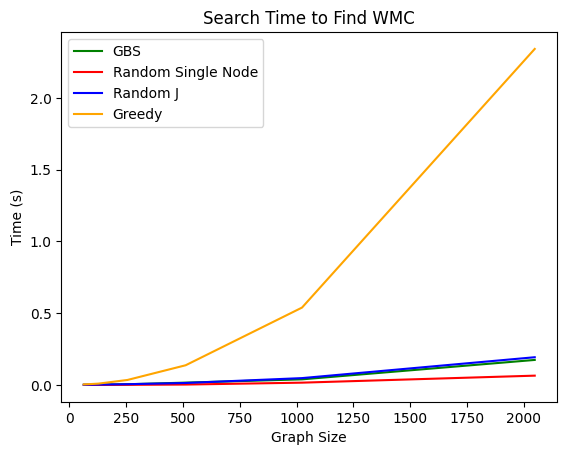}
\caption{Plot of time versus graph size for the weighted maximum clique problem.}
\label{fig:plantedWMCInit}
\end{figure}

In Figure \ref{fig:wmdormalizedlogscale}, we once again observe that the GBS strategy consistently provides highly weighted solutions, with the difference being increasingly evident at higher graph sizes. In Figure \ref{fig:WMCsamples} we observe consistency of the GBS strategy, being able to produce its high weighted solution in significantly less samples.
From this we can conclude that the weighted max clique task heavily benefits from the GBS strategy over other randomized seed algorithms. Moreover, the benefits of this strategy are more evident at larger node counts indicating great potential for scalability. 

\subsection{Graph Classification}

Finally, we explored the application of GBS to the task of graph classification.  The previous applications were optimization problems, with the intent of finding a particular structure with some highest score. However, the problem of graph classification asks a more holistic question regarding the properties of the graph itself rather than the structures it contains.  The definition of the graph classification problem is as follows: Given a set of graphs, each with a categorical label, predict the label associated with a graph that has not been encountered before. In our case, the goal is to effectively train a machine learning model to classify the graphs.

This task simplifies to the smaller subtask of quantifying the similarity between two graphs. In particular, given two graphs $G_1$ and $G_2$, the goal is to calculate a score that quantifies the similarity between $G_1$ and $G_2$.  If two graphs are more similar to each other, they are more likely to have the same label and belong to the same category. A special case of this is graph isomorphism: if $G_1$ and $G_2$ are equivalent up to a permutation of nodes, then the similarity score is maximized.  The general approach to computing graph similarity is as follows: First, we must introduce a feature map that transforms a graph to a multidimensional vector that embeds information about the graph. Then, a machine learning kernel must quantify the distance between these two feature vectors, which correspond to $G_1$ and $G_2$.

A GBS-based strategy can be applied to graph isomorphism and classification by using the device's sampling as the mapping. Since sampling from a GBS device will sample from the dense regions of the encoded graph, when comparing the samples of two different graphs we are essentially comparing their defining characteristics.  In the ideal case, two graphs are equivalent up to a permutation of nodes (i.e., they are isomorphic). Over many samples, they would have the same sampling distribution, up to a permutation.  In order to efficiently compute feature vectors, rather than basing comparisons off of individual measurements, we will use a coarse-grained sampling strategy. This involves maintaining groups of samples, and for threshold-based GBS, there are two strategies.

The first strategy, called count binning, involves grouping samples based on the number of total detector clicks measured in the experiment. The second strategy, called detector binning, involves grouping samples based on the particular detector that clicked. With both strategies, after taking multiple samples a cumulative frequency vector of size $M$ is created, where $M$ is the number of modes in the GBS device as well as the number of nodes in the graph. While these vectors can be compared directly, they can also be fed into a support vector machine (SVM) as feature vectors, allowing machine learning techniques to explore and compare various substructures of a particular graph. One advantage of these feature vectors is that their size scales linearly with the number of modes, unlike PNR-based strategies explored in the literature \cite{Bradler2021}.  This reduces the number of samples needed to approximate the feature vector within a certain error bound. In particular, it is stated in \cite{Anteneh2023} that it takes $\mathcal{O}(\frac{M+\log(\frac{1}{\delta})}{\epsilon^2})$ samples to compute the count-binning feature vector such that the sum of the absolute values of the errors of its entries is greater than $\epsilon$ with a probability no greater than $\delta$. 

Using these strategies, we trained a support vector machine using the RBF kernel to classify a variety of data sets featuring different graph types \cite{Morris+2020}. 
We used the same parameters as in \cite{Anteneh2023} in training our model and generating samples, except that the maximum number of detector clicks was unbounded as this is not a parameter in the TGBS simulation. 
Specifically, the size of the graphs ranges from 6 to 25 nodes, and node labels, node attributes and edge labels are ignored. A total of 6000 TGBS samples are taken from each graph. Additionally, we use $\overline{n} = 5$ as the mean photon number. Finally, the model accuracy is obtained using double 10-fold cross-validation 10 times, and the $C$ hyperparameter is obtained through a grid search over the values $[10^{-4}, 10^{-3}, ..., 10^2, 10^3]$. Results are shown in Table I. 
\begin{center}
\begin{table}[ht]
\begin{tabular}{ |p{1.7cm}|p{3cm}|p{3cm}|  }
 \hline
 \multicolumn{3}{|p{7.7cm}|}{Mean Accuracy for datasets of graphs with 6-25 nodes using nested 10-fold cross validation. Generated using 6000 samples, mean photon number = 5.} \\
 \hline
 Dataset & Count Binning Mean Accuracy \% $\pm$ std & Detector Binning Mean Accuracy \% $\pm$  std  \\
 \hline
 AIDS   & $99.65 \pm 0.00$ & $97.95 \pm 0.11$ \\
 BZR\_MD &   $63.55 \pm 0.51$  & $64.16 \pm 0.47$ \\
 COX2\_MD &  $52.00 \pm 1.47$ &  $50.63 \pm 3.03$\\
 IMDB-BINARY&   $66.42 \pm 0.40$ & $66.83 \pm 0.67 $\\
 MUTAG& $82.17 \pm 0.46$ & $77.37 \pm 1.54$\\
 NCI1&  $53.79 \pm 2.32$  & $52.94 \pm 1.75$ \\
 PROTEINS& $62.97 \pm 0.40$  & $67.34 \pm 1.14$\\
 PTC\_FM& $58.57 \pm 2.16$ & $52.67 \pm 3.72$ \\
 \hline
\end{tabular}\\
\caption{Count binning and detector binning mean accuracies with TGBS on datasets with graphs of 6-25 nodes.}
\label{tab: SmallDatasetGBSAccuracy}
\end{table}
\end{center}




Additionally, we benchmarked the GBS count binning and detector binning SVM approaches for larger datasets to demonstrate scalability and offer comparisons against common classical kernalized classifiers such as Random Walk, implemented by the GraKeL library \cite{JMLR:v21:18-370}. These datasets are binary (2 possible labels for a graph) with the exception that the COLLAB and Synthie datasets consist of 3 and 4 classes of graphs respectively. As with the smaller graphs, we ignored node labels, node attributes and edge labels as these cannot be embedded into a GBS device; however, other classical algorithms can certainly take advantage of these for better classification depending on the dataset. The time complexity for count binning and detector binning is $\mathcal{O}(M^3)$, which is the same as the Random Walk kernel and theoretically advantageous over common classical kernels including the Shortest-Path Kernel and Graphlet Sampling Kernel (for graphs with unbounded node degree). Now, to account for imbalanced datasets, we utilize a the balanced accuracy metric to evaluate performance. This score is the average value of the recall achieved on each class within a dataset. The recall score can be computed as
\begin{equation}
    \textnormal{Balanced Accuracy} = \frac{1}{K}\sum_{i=1}^{K} \frac{\textnormal{True Positives for class }i}{\textnormal{Number of instances of class }i} 
    \label{eq: balanced accuracy score}
\end{equation}
where $K$ is the number of classes in the dataset. Like with the graphs ranging from 6-25 nodes, the balanced accuracy is evaluated using nested 10-fold cross validation, and the hyperparameter $C$ is obtained through a grid search. Finally, the machine learning process (after obtaining samples) was given 24 hours to run before timing out.
Results are shown in Table II. 

As Table II shows, the GBS kernels were able to achieve comparable accuracy on most datasets and even significantly exceed the Random Walk kernel on certain datasets. For example, the detector binning kernel outperformed Random Walk on SYNTHETIC-new, and both GBS kernels outperformed Random Walk on DD. However, Random Walk achieved a balanced accuracy greater than 10 percentage points than the detector binning kernel and greater than 15 percentage points than the count binning kernel on the COLLAB dataset.

\begin{center}
\begin{table}
\begin{tabular}{ |p{1.6cm}|p{1.1cm}|p{1.5cm}|p{1.5cm}|p{1.5cm}|  }
 \hline
 \multicolumn{5}{|p{8.8cm}|}{Mean Balanced Accuracy for graphs up to 500 nodes using GBS Count Binning, GBS Detector Binning and Random Walk Kernels, calculated using nested 10-fold cross validation. A maximum of 200000 samples were used.} \\
 \hline
 Dataset & Nodes [min,max] & Count Binning Balanced Accuracy \% $\pm$ std & Detector Binning Balanced Accuracy \% $\pm$ std & Random Walk Balanced Accuracy \% $\pm$ std \\
 \hline
 Synthie   & [0,200] & $ 47.32 \pm 0.59$ & $48.51 \pm 1.92$ & $49.79 \pm 0.07$ \\
 SYNTHETIC-new & 100 & $49.43 \pm 2.72$ & $62.33 \pm 1.15$   & $52.67 \pm 0.00$ \\
 COLLAB & [100,200]  & $70.25 \pm 1.24$ &$75.05 \pm 1.62$ &  $85.43 \pm 1.49$\\
 DD    & [100,200]  & $57.78 \pm 1.01$ & $54.31 \pm 1.41 $ &  $49.06 \pm 1.15$ \\
 REDDIT-BINARY & [400,500]  & $76.17 \pm 1.02$ & $79.05 \pm 1.25$ & $81.98 \pm 0.66 $\\

 \hline
\end{tabular}\\
\caption{Count binning, detector binning, and random walk mean balanced accuracies for datasets with graphs of up to 500 nodes.}
\label{tab: LargeDatasetGBSBalancedAccuracy}
 
\end{table}
\end{center}

\begin{center}
\begin{table}
\begin{tabular}{ |p{1.6cm}|p{1.1cm}|p{1.5cm}|p{1.5cm}|p{1.4cm}|  }
 \hline
 \multicolumn{5}{|p{8.8cm}|}{Graph classification kernel approximate runtimes for graphs to 500 nodes using GBS Count Binning, GBS Detector Binning, and Random Walk Kernels.} \\
 \hline
 Dataset & Nodes [min,max] & Count Binning Runtime (seconds) & Detector Binning Run Time (seconds)  & Random Walk Runtime (seconds) \\
 \hline
 Synthie   & [0,200] & $610$ & $643$ & $1216$ \\
 SYNTHETIC-new & 100 & $507$ & $499$ & $616$\\
 COLLAB & [100,200]  & $1169$ & $1145$ & $8800$\\
 DD    & [100,200]  & $815$ & $818$ & $2656$ \\
 REDDIT-BINARY & [400,500]  & $2364$ & $2368$ & $55757$\\

 \hline
\end{tabular}\\
\caption{Count binning, detector binning, and random walk start-to-finish runtimes for datasets with graphs of up to 500 nodes.}
\label{tab: LargeDatasetGBSBalancedAccuracy}
 
\end{table}
\end{center}

\subsection{GBS Overhead Calculations}
As per our results, we witness that in implementation, GBS seeded algorithms follow theoretical claims do indeed outperform classically seeded algorithms in terms of solution quality and convergence time. However, for GBS to have true utility over these algorithms, the entire process of GBS must be comparable in time to the classically seeded ones. This necessitates the inclusion of the total runtime of GBS, most importantly the encoding of the graph problem. As stated before, this TA decomposition has a time complexity of $\mathcal{O}(M^3)$ which clearly dominates the classical overhead for random heuristics requiring $\mathcal{O}(j)$ and for the greedy peeling strategy which requires $\mathcal{O}(1)$ time.

\begin{figure}[ht]
\includegraphics[width=8cm]{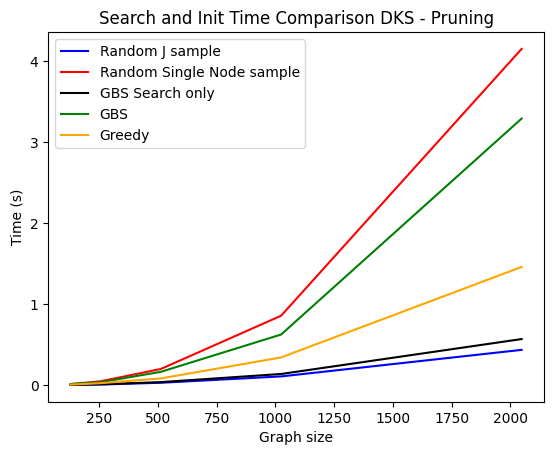}
\caption{Plot of time including decomposition versus graph size for densest k-subgraph solutions.}
\label{fig:dksinittimes}
\end{figure}

\begin{figure}[ht]
\includegraphics[width=8cm]{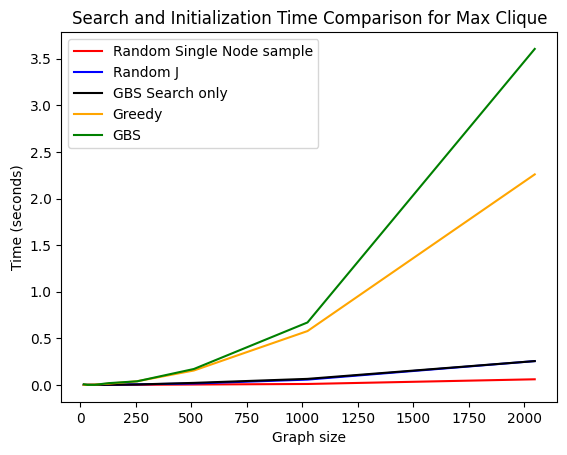}
\caption{Plot of time including decomposition versus graph size for max clique solutions.}
\label{fig:MCinittimes}
\end{figure}

\begin{figure}[ht]
\includegraphics[width=8cm]{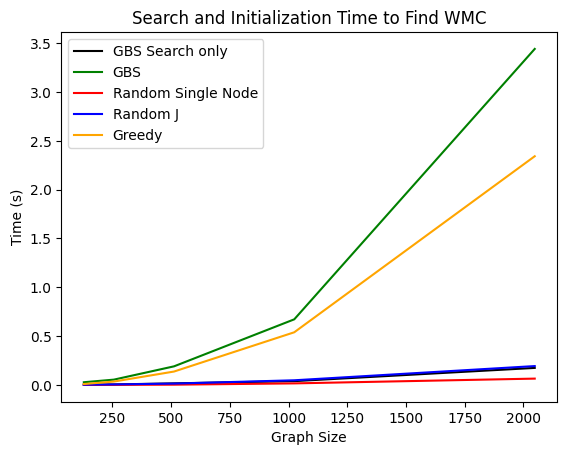}
\caption{Plot of time including decomposition versus graph size for weighted max clique solutions.}
\label{fig:WeightedMCinittimes}
\end{figure}

We observe from Figures \ref{fig:dksinittimes}, \ref{fig:MCinittimes}, and \ref{fig:WeightedMCinittimes} that when including the graph encoding time, the GBS time greatly spikes and in some cases, becomes 100 times that of the competing seeded strategies.  Furthermore, we note that the improvement in solution quality is marginal in comparison. 

Thus, when comparing the total time taken by GBS for the non-kernel based graph problems, it becomes apparent that practical utility is difficult to justify for generic random graphs. This issue is not unique to our TGBS algorithm --- the TA decomposition process is a requirement for embedding graph problems even into hardware implementations of GBS. However, there is something to be said for graphs which are more structured.
Current research aims to speed up the TA decomposition process, taking advantage of classical low-rank approximations of singular value decomposition \cite{doi:10.1137/090771806}. This could prove especially useful in graphs which contain prominent underlying structures, where patterns within subgraphs correlate to patterns on the full graph. as opposed to our Erd\"{o}s–R\'{e}nyi models.

\section{Conclusion}
\label{sec:Con}
In this work we have gauged the practical advantage gained by GBS through a TGBS device. While numerical demonstrations have been shown in the past, we experimentally compare against current classical seeded methods to verify the theoretical claims and determine the practical utility of the GBS based method.

Critically, while the GBS solution aims to provide a global approximate solution for these tasks, it does so whilst not having to brute force the classically intractable problem. The TGBS algorithm, which serves as a lower bound for GBS samples in comparison to an exact device, provides this improvement utilizing a polynomial time algorithm including its overhead costs.

However, it is important to note that the biggest obstacle for the GBS solution in achieving universal practical utility is the TA decomposition runtime. Although approximation techniques do exist and are currently being researched, they do not work well for generic random graphs that contain few structural patterns. This reduces the current applicability of GBS to more structured problems and/or problems for which current solvers exceed a $\mathcal{O}(M^3)$ runtime. 

Thus, further work is required to fully assess the utility of GBS based algorithms but there exist promising avenues. As demonstrated in our work graph classification tasks are able to benefit since the decomposition time for GBS is dominated by the other overheads required in the machine learning process.

Another promising pattern of problems is the Maximum Weighted Clique, as the GBS algorithm demonstrates much higher quality solutions consistently across various sizes of graphs. Additionally, the problem encoding process involves the $\Omega$ hyperparameter, which allows us to adjust the focus to the weighted portions of the graph and fine-tune our control over particular structures of the graph. This could prove to be especially useful when more context on the distribution of weights or edges on the graph is given, such as with graphs regarding communication networks or roads between cities.

A known interesting problem which can be solved through the Maximum Weighted Clique problem is that of finding RNA secondary structures. While the best classical methods require a large amount of overhead in the form of support vector machines or even deep neural networks to take into account advanced 3-D structures such as pseudoknots, a mapping to MWC allows GBS to provide an approximate solution in a mere fraction of the time. This is especially important when there are multiple strands of RNA involved, where the graphic representation would span thousands and even millions of nodes.



\section*{Acknowledgments}

This work was supported by the Office of Naval Research under Grant No.\ {N00014-23-1-2115} and by Applied Research Laboratories, The University of Texas at Austin, under an internal research and development award.

\ifoverleaf
\bibliographystyle{unsrt}
\else
\bibliographystyle{IEEEtran}
\fi
\bibliography{refs}

\end{document}